\documentclass[aps,prb,showpacs,twocolumn,superscriptaddress]{revtex4}

\usepackage{graphicx}
\usepackage[latin1]{inputenc}

\begin{document}

\newcommand{\si}{sample~I}
\newcommand{\sii}{sample~II}
\newcommand{\siii}{sample~III}

\title{Influence of topography and Co domain walls on the magnetization reversal of the FeNi layer in FeNi/Al$_2$O$_3$/Co
magnetic tunnel junctions}
\author{F.~Romanens}
\author{J.~Vogel}
\affiliation{Laboratoire Louis N\'{e}el, CNRS, 25 avenue des
Martyrs, BP~166, F-38042 Grenoble cedex~9, France}
\author{W.~Kuch}
\author{K.~Fukumoto}
\affiliation{Institut f\"{u}r Experimentalphysik, Freie Universität
Berlin, Arnimallee 14, D-14195 Berlin, Germany}
\author{J.~Camarero}
\affiliation{Dpto. F\'{i}sica de la Materia Condensada, Universidad
Aut\'{o}noma de Madrid, E-28049 Madrid, Spain.}
\author{S.~Pizzini}
\author{M.~Bonfim}
\affiliation{Laboratoire Louis N\'{e}el, CNRS, 25 avenue des
Martyrs, BP~166, F-38042 Grenoble cedex~9, France}
\author{F.~Petroff}
\affiliation{Unit\'{e} Mixte de Physique CNRS/Thales, Route
d\'{e}partementale 128, F-91767 Palaiseau cedex, France}

\date{October 5, 2006}

\begin{abstract}
We have studied the magnetization reversal dynamics of
FeNi/Al$_2$O$_3$/Co magnetic tunnel junctions deposited on
step-bunched Si substrates using magneto-optical Kerr effect and
time-resolved x-ray photoelectron emission microscopy combined with
x-ray magnetic circular dichroism (XMCD-PEEM). Different reversal
mechanisms have been found depending on the substrate miscut angle.
Larger terraces (smaller miscut angles) lead to a higher nucleation
density and stronger domain wall pinning. The width of domain walls
with respect to the size of the terraces seems to play an important
role in the reversal. We used the element selectivity of XMCD-PEEM
to reveal the strong influence of the stray field of domain walls in
the hard magnetic layer on the magnetic switching of the soft
magnetic layer.
\end{abstract}


\maketitle

\section{Introduction}
Magnetic tunnel junctions and spin valves are extensively used in
magnetic storage devices, and are also interesting from a
fundamental point of view. Their active part, composed of two
ferromagnetic (FM) materials with different coercivities separated
by a non-magnetic (NM) spacer layer, presents a number of
interesting phenomena like giant
magnetoresistance\cite{Baibich1988,Binasch1989} and spin transfer
torque.\cite{Myers1999,Weber2001} In these trilayer systems,
magnetostatic effects can strongly influence the magnetization
reversal. For instance, an interaction between the two magnetic
layers can be induced by correlated roughness at the FM/NM
interfaces. This roughness induces magnetic charges at the
interfaces, and the interaction between them leads to the so-called
``orange-peel'' coupling,\cite{Neel1962} favoring  a parallel
alignment of the magnetization of the two ferromagnetic layers. The
roughness also influences the domain wall pinning and therefore the
magnetization reversal. In a previous paper,\cite{Pennec2004} we
have revealed the effect of modulated roughness, induced by
deposition on step-bunched Si substrates, on the magnetization
reversal and coupling in magnetic trilayers. Steps with an
orientation perpendicular to the easy magnetization axis are at the
origin of a strongly localized orange-peel
coupling.\cite{Pennec2004} On the other hand, steps parallel to the
easy magnetization axis induce strong demagnetizing effects when
domain walls are located on these steps. This causes a pinning of
the domain walls that hinders reversal by domain wall propagation.
One might expect this effect to be particularly important when the
difference in energy between domain walls situated on and situated
between steps is large. To test the influence of the topography on
the magnetization reversal, we have investigated FM/NM/FM trilayers
deposited on step-bunched substrates with different miscut angles,
leading to different widths of the terraces and thus to different
distances between steps. We have used magneto-optical Kerr effect
and time-resolved x-ray magnetic circular dichroism combined with
photoemission electron microscopy (XMCD-PEEM) to obtain both a
global and a detailed microscopic view of the influence of the
substrate-induced layer topography on the magnetization reversal.
These measurements show that in the samples with the largest
terraces nucleation of reversed domains is easier while the pinning
of domain walls is stronger than in samples with smaller terraces.
In order to explain this we suggest that  the width of domain walls
with respect to the average width of terraces has to be taken into
account.

Apart from magnetostatic effects induced by layer topography, stray
fields from inhomogeneously magnetized regions in one of the layers
can also influence the static and dynamic magnetic properties of the
other layer. For example, domain walls in the soft magnetic layer
can create stray fields that are large enough to influence the
magnetization of the hard magnetic layer in soft~FM/NM/hard~FM
trilayers, as shown by Thomas \emph{et al}.\cite{Thomas2000} More
recently, several authors have shown direct evidence of the effect
of a domain wall in one layer on the static magnetic configuration
of the other
layer.\cite{Schafer2002,Kuch2003,Christoph2004,Wiebel2005} In a
recent paper,\cite{Vogel2005} we have used the element selectivity
of XMCD-PEEM to study independently the magnetization of both
magnetic layers in FeNi/Al$_2$O$_3$/Co trilayers, showing that
domain walls in the hard magnetic layer locally decrease the
nucleation barrier for the switching of the soft magnetic layer. In
this paper, we have extended this study in order to obtain
information on the relative influence of topography and domain wall
stray fields on the nanosecond magnetization reversal in spin-valves
and magnetic tunnel junctions.

\section{Experimental methods}
The samples studied were  Fe$_{20}$Ni$_{80}$(4~nm)/ Al$_2$O$_3$
(2.6~nm)/ Co(7~nm)/ CoO(3~nm) magnetic tunnel junctions  deposited
on Si(111) substrates by RF sputtering. The CoO layer was used to
increase the coercive field of the Co layer. In fast dynamic
measurements, coercivities increase and the field range over which
the magnetization reverses (magnetization transition) is broader
than in quasi-static conditions.\cite{Pennec2004} A large difference
in quasi-static coercivity between the two magnetic layers is
therefore needed in order to allow the FeNi magnetization to be
switched without changing the Co magnetization in our fast dynamic
measurements.

Three different samples (named samples I, II, and III) were
deposited on substrates with miscut angles along the [11$\overline
2$] direction of 4$^\circ$, 6$^\circ$, and 8$^\circ$ respectively.
After annealing, all the substrates present a step-bunched surface
with ellipsoidal terraces.\cite{SussiauPhD} Transmission electron
microscopy images show that these topographic features are well
reproduced by the subsequently deposited layers.\cite{Pennec2004}
The size of the terraces and the height of the steps depend on the
miscut angle (for a miscut angle of 4, 6 and 8$^\circ$, the average
terrace width is about 60, 40, and 20~nm, and the step height is
about 5, 4, and 3~nm respectively). The steps make an angle of about
60$^\circ$ with respect to the normal of the surface (see Fig. 1 in
Ref~\cite{Pennec2004}). This topography induces a uniaxial in-plane
magnetic anisotropy in all three samples, with the easy axis
parallel to the long axis of the terraces.

We have studied the magnetization reversal of these samples using
Kerr magnetometry and XMCD-PEEM. Longitudinal Kerr effect
measurements were performed by illuminating the samples with a
linearly polarised He-Ne laser beam. The polarization rotation of
the reflected light due to the Kerr effect is detected by a
combination of a Wollaston prism and a pair of photodiodes. For
``slow'' magnetization reversal (for a field sweep rate $dH/dt$
below 10~T/s) the field was applied using a ferrite electromagnet.
For ``fast'' magnetization reversal, ``strip-line'' coils made of a
hairpin-shaped copper ribbon into which the samples were inserted
was used. A window was opened in the top part of the coil in order
to allow illumination.\cite{Vogel2004} These coils, in combination
with a fast current generator, allow magnetic fields up to 20~mT to
be generated, with a rise-time of the order of 10~ns. The error bar
on the field values obtained with this coil, given in the text and
figures, is estimated to be about 10\%. Kerr magnetometry allows
macroscopic magnetization reversal to be measured for a wide range
of time scales (from quasi-static to the nanosecond timescale).
However it does not give local information about the magnetic
switching as it measures the sample's magnetization integrated over
the laser spot size, which was about 100~$\mu$m in our measurements.

In order to obtain local information, we have performed magnetic
imaging using time resolved XMCD-PEEM with the coil mentioned above.
These measurements were carried out at beamlines UE56-2 and UE52 of
the BESSY synchrotron in Berlin. The sample is illuminated with
circularly polarized x-rays. The XMCD mechanism causes the x-ray
absorption of the sample to depend on the relative orientation of
the local magnetization and the helicity vector of the
x-rays.\cite{Stohr1999,Wende2004} Absorption of x-rays creates
photo-emitted electrons with an intensity that is proportional to
the local absorption. These electrons are collected by an electron
microscope and projected on a CCD camera, therefore allowing to form
an image of the sample in which the intensity represents the
projection of the local magnetization on the direction of the x-ray
propagation. \cite{Schneider2002,Kuch2004} In order to enhance the
magnetic contrast and subtract the topographic component, the final
image is the difference between images taken with right and left
circularly polarized x-rays. One powerful feature of this technique
is its element selectivity, i.e. by tuning the photon energy to the
Fe L$_3$ or Co L$_3$ absorption edge, the permalloy or cobalt layer
magnetization can be probed independently. When applying a magnetic
field on the sample, the trajectory of the photo-emitted electrons
is changed, leading to a shift of the image. In order to correct
this shift, we have translated the images in
Fig.~\ref{fig:PEEM_mis4deg}-\ref{fig:PEEM_mis8deg} so that the same
region of the sample is shown on each image independently on the
applied field.

Temporal resolution was obtained by exploiting the time-structure of
synchrotron radiation. In the single bunch operation mode of BESSY,
photon pulses are emitted with a repetition rate of 1.25~MHz. The
temporal resolution of our measurements, defined by the width of the
photon pulses ($\sim$70 ps) and electronic jitter, was better than
100 ps. Measurements were performed in pump-probe mode, by
synchronizing the photon pulses with the applied field
pulses.\cite{Bonfim2001,Vogel2003} By tuning the delay between the
photon pulse and the magnetic field, magnetization reversal was
studied as a function of time during the magnetic pulse.

\section{Magnetization reversal}

\begin{figure}[ht!]
\includegraphics[width=7.5cm]{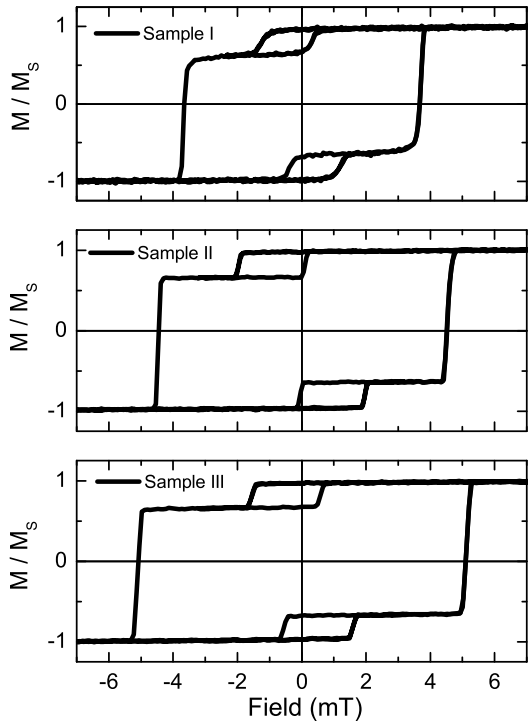}
\caption{\label{fig:cycles} Quasistatic hysteresis loops and minor
loops for magnetic tunnel junctions deposited on step-bunched Si
with a miscut angle of 4$^o$ (top), 6$^o$~(middle) and
8$^o$~(bottom). The hysteresis loops are taken with the magnetic
field applied along the easy magnetization axis (parallel to the
long axis of the substrate terraces).}
\end{figure}

The quasistatic hysteresis loops for the three samples obtained
using longitudinal Kerr effect measurements are shown in
Fig.~\ref{fig:cycles}. The hysteresis loops were recorded with the
magnetic field applied along the long axes of the terraces,
corresponding to the easy magnetization axis. Because of the
presence of the CoO layer, the coercive field of the Co layer is
much larger than the coercive field of the FeNi layer, thus leading
to well defined FeNi minor loops. These FeNi minor loops are not
centered on zero-field due to the so-called Néel ``orange peel''
coupling between the two magnetic layers. This magnetostatic
coupling, due to the presence of steps between terraces, has been
previously studied, and its manifestation depends on the switching
mode of magnetization.\cite{Pennec2004}

The roundness of the FeNi minor loop for \si{} indicates that domain
nucleation plays an important role in the reversal of this sample,
and that domain walls are strongly pinned. The hysteresis loops of
\sii{} and III are square shaped. This indicates that the coercive
fields of these samples are determined by the field needed to
nucleate domains and that once a domain is nucleated, magnetization
reversal takes place by fast propagation of domain walls. The
transitions in the minor loop of sample III are more tilted than for
\sii{}. The loops in Fig.~\ref{fig:cycles} are averages over several
(some tens) of single hysteresis loops. For sample \siii{}, we
observed a distribution of reversal fields, probably due to
different nucleation positions and therefore different times/fields
for which a domain wall crosses the laser spot. In the averaged
loops, the FeNi magnetization transition therefore appears less
square.

\begin{figure}[ht!]
\includegraphics[width=8.4cm]{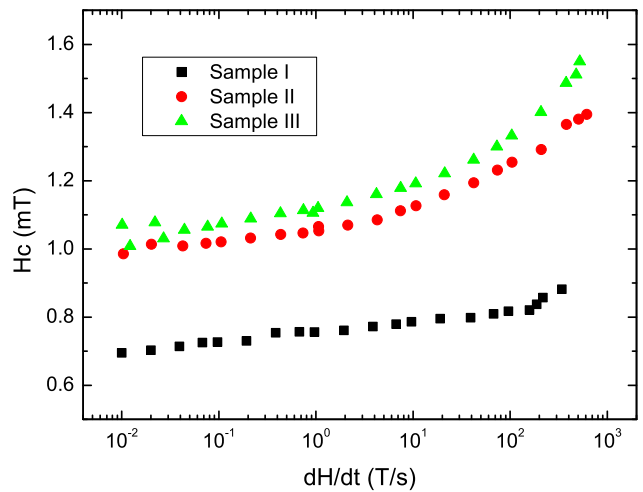}
\caption{\label{fig:dhdt} Coercive field of the Fe$_{20}$Ni$_{80}$
layer as a function of the applied field sweep rate $dH/dt$.}
\end{figure}

We have measured hysteresis loops with field sweep rates ranging
from 10~mT/s to 1~kT/s. The permalloy coercive field as a function
of $dH/dt$ is plotted in Fig.~\ref{fig:dhdt}. The coercive field was
taken as the average of the positive and negative reversal fields of
the minor hysteresis loops. When increasing the applied field sweep
rate, dynamic effects appear such as an increase of the coercive
field and a broadening of the magnetization transitions.

For the lowest field sweep rates, the increase of the permalloy
coercive field is quite linear with the logarithm of the applied
field sweep rate as expected from theory.\cite{Raquet1995} In a
previous work,\cite{Camarero2001} a clear deviation from this linear
regime at fast sweep rates was interpreted as a transition from a
propagative reversal at low sweep rates to a more nucleative
reversal at fast timescales, as suggested by
Raquet.\cite{Raquet1996} In our case, for \si{} a deviation from
linearity is well observed at a sweep rate of 100~T/s, but the shape
of the quasistatic hysteresis loop indicates that even at very low
sweep rates nucleation already plays an important role in the
magnetization reversal. The observed transition can, however, be due
to a sudden increase of the nucleation density at high field sweep
rates. For samples II and III, no clear transition is visible in the
investigated range of field sweep rates. This means that either
there is no critical field at which a sudden transition from domain
wall propagation to nucleation dominated reversal takes place, or
that this critical field falls outside the measured range. However,
the interpretation of the $H_c(dH/dt)$ data is difficult and not
clear enough to conclude on the mechanisms governing the
magnetization reversal.

\begin{figure}[t!]
\includegraphics[width=8.4cm]{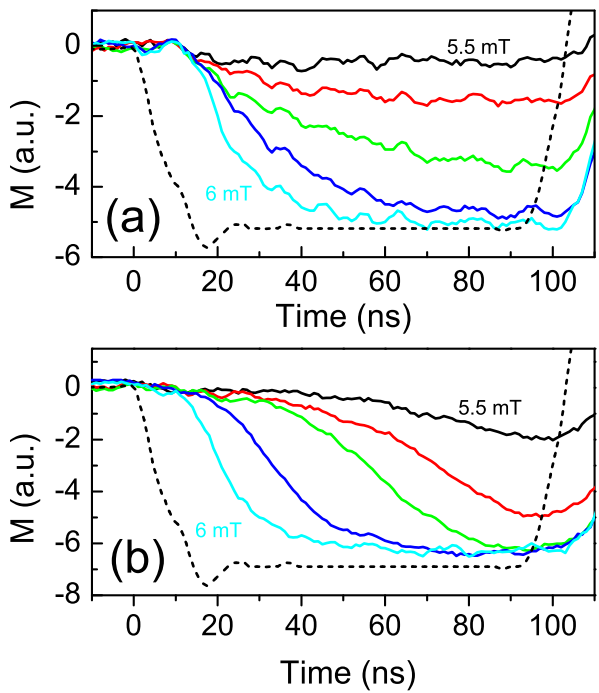}
\caption{\label{fig:relax} Fast magnetization relaxation for \sii{}
(a) and \siii{} (b). The magnetization decay is plotted in solid
lines for different applied field values between 5.5 and 6 mT. The
shape of the magnetic field pulse for a pulse with an amplitude of
6~mT is shown as the dashed line. A synchronous noise can be seen in
(a). This noise is due to electromagnetic interference created by
the current generator and the coil.}
\end{figure}

For a better understanding of these mechanisms, we have carried out
measurements of magnetization relaxation at the nanosecond
timescale. Magnetization relaxation has been used in
Co/Pt\cite{Romanens2005a} and other systems where domain wall
pinning is important. The advantage of relaxation measurements with
respect to measurements with varying field sweep rates is that the
magnetic field is fixed during the magnetization reversal. In planar
thin films, the reversal is very fast, thus relaxation measurements
on our samples were possible only for very fast risetimes of the
magnetic field. In our magnetization relaxation measurements, the
samples were initially saturated in the positive direction using a
quasi-static magnetic field. A negative field pulse was then applied
at $t=0$ along the easy magnetization axis (see dotted lines in
Fig.~\ref{fig:relax}), with an amplitude sufficient to reverse the
permalloy magnetization but low enough so that the cobalt
magnetization did not change. After the 100~ns long negative field
pulse, a positive pulse (not shown in Fig.~\ref{fig:relax}) was
applied in order to saturate the permalloy magnetization again in
the positive direction. The field was applied with a rise time of
about 10~ns, and the amplitude of the field was small enough for the
reversal to take place mainly during the plateau in which the field
was constant. Due to the small barrier for nucleation of reversed
domains in \si{}, the magnetization of this sample always started
reversing before the field plateau. In Fig.~\ref{fig:relax}, only
the relaxation curves for \sii{} and \siii{} are therefore shown.
Each curve in Fig.~\ref{fig:relax} is an average over about 500
relaxation curves, to increase the signal to noise ratio.

It can be seen in Fig.~\ref{fig:relax} that the field needed to
reverse the permalloy layer (against the direction of the Co
magnetization) in 100~ns is about 5.7-5.9 mT. This is a factor 2-3
higher than for the quasi-static measurements. This value is in
agreement with the fast increase of coercivity expected from an
extrapolation of the curves in Fig.~\ref{fig:dhdt} (5-6 mT in 100 ns
gives an average dH/dt of 5-6$\times$10$^4$ T/s). A similar behavior
has been observed for Co/Pt systems\cite{Moritz2004}.

Oscillations due to precessional-like magnetization reversal have
been observed in several micrometer-sized thin film
structures\cite{Schumacher2003,Bailey2004,Choi2005,Krasyuk2005}. In
these structures, the zero-field ground state is a flux-closure
domain state due to finite-size and demagnetizing field effects,
which also have an important influence on the magnetization
dynamics. They play no significant role in the magnetization
dynamics of our samples, consisting of in-plane magnetized
continuous films for which homogeneous magnetization along the easy
axis gives the lowest energy. In that case, the magnetization
reversal occurs through incoherent nucleation and propagation
processes that are not expected to give rise to oscillations in the
magnetization behavior. The oscillations observed in the relaxation
curves of Fig.~\ref{fig:relax}, especially for sample II, are due to
synchronous noise, caused by the electromagnetic interference
between the pulsed current generator and the magnetic coil.

Both sets of relaxation curves can be understood by the model
initially developed by Fatuzzo for ferroelectric
materials\cite{Fatuzzo1962} and later adapted to ferromagnetism by
Labrune.\cite{Labrune1989} According to this model, the relaxation
curve is exponential in the case of nucleation-dominated reversal,
and S-shaped in the case of propagation-dominated reversal. A clear
difference between the behavior of the two samples is observed in
these fast relaxation measurements. Sample II has a quasi
exponential relaxation, indicating more nucleation than in \siii{}
which has an S-shaped relaxation curve. Even if there are no
significant differences in the hysteresis loops nor in the
$Hc(dH/dt)$ data, these relaxation measurements indicate that, at
these short timescales, the reversal is more propagative in \siii{}
than in \sii{}.

In summary, magnetometry measurements indicate that nucleation plays
an important role in the reversal of \si{}, whereas domain wall
propagation dominates the reversal of \sii{} and III in quasistatic
conditions. During faster reversal, the nucleation becomes more
important in the magnetization reversal of sample II, indicating a
change of regime compared to quasistatic conditions. For sample III,
domain wall propagation is still the process dominating the reversal
even at short timescales.

\begin{figure}[ht!]
\includegraphics*[bb= 0 631 234 841]{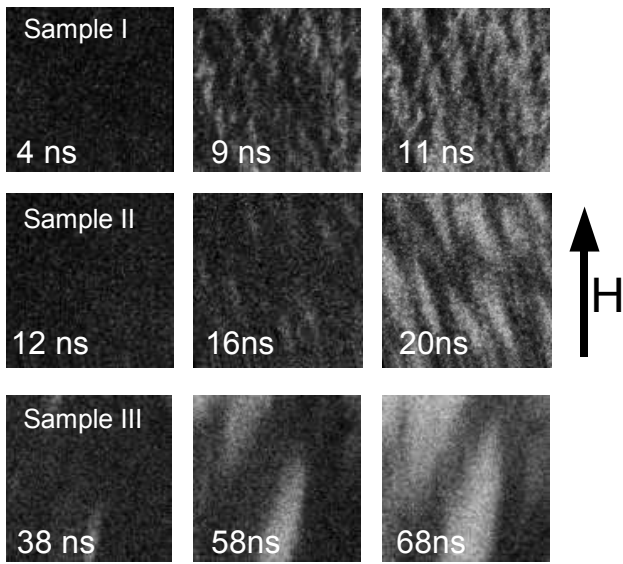}
\caption{\label{fig:PEEM_Co_black} Magnetization reversal of the
FeNi layer in the different FeNi/Al$_2$O$_3$/Co trilayers. The
applied magnetic field is about 6.5~mT for \si{} (top row) and
\sii{} (middle row), and about 6~mT for \siii{} (bottom row). The
projection of the x-ray incidence direction on the sample surface is
pointing up in the images (parallel to the arrow) and is parallel
(anti-parallel) to the direction of the field for positive
(negative) pulses. The magnetization direction is in the plane of
the layers and points up (parallel to the arrow) for black domains,
and down for white domains. The time delay with respect to the
beginning of the (negative) magnetic pulse is indicated in the
images. The field of view is about 50~$\mu$m for all the images. }
\end{figure}

In order to confirm these conclusions with microscopic measurements,
we have performed time-resolved XMCD-PEEM measurements. Figure
\ref{fig:PEEM_Co_black} shows the evolution of the domain structure
of the permalloy layer during the application of a magnetic pulse
applied along the easy axis of magnetization. The trilayer is
initially saturated in one direction, giving a black image. In this
condition, permalloy and cobalt magnetizations are parallel. At time
$t=0$ the magnetic pulse triggers magnetization reversal of the
permalloy layer toward antiparallel alignment with Co. The reversal
occurs by nucleation of reversed domains, which appear as white
contrast in Fig.\ref{fig:PEEM_Co_black}, and a subsequent
propagation of domain walls.

The images clearly show that for similar field values the number of
nucleated domains strongly decreases going from \si{} to \siii{}.
Actually, the main part of the reversal in \si{} takes place during
the risetime of the field pulse, while the dynamics of the domain
walls at the plateau is very slow. In \siii{} the density of
nucleated domains is very small, and the images for this sample are
quite blurred. Note that the images are acquired in a pump-probe
mode, averaged over about $10^8$ magnetic and photon pulses. The
blurred images of \siii{} therefore indicate that the domain wall
motion that dominates the reversal in this sample is less
reproducible than domain nucleation. A wide distribution of
nucleation barrier energies seems to exist in these samples, which
is also indicated by the strong increase of the number of nucleated
domains upon increasing pulse height.\cite{Fukumoto2006} The
nucleation is therefore mainly field induced and thermal activation
plays a minor role.

We conclude that both macroscopic magnetometry and microscopic
magnetic imaging measurements indicate that the reversal mechanism
for the FeNi/Al$_2$O$_3$/Co trilayers depends on their topography.
While the nucleation density decreases going from \si{} (larger
terraces) to \siii{} (narrower terraces), the domain wall mobility
increases and domain wall pinning therefore decreases.

We propose the following explanation for these results. The energy
of a domain wall will depend on whether it is on a terrace or on a
step. In these thin films (thickness $<$10 nm), domain walls are of
the N\'{e}el-type.\cite{Hubertbook} If located on a step parallel to
the easy magnetization axis, the magnetization in the center of the
domain wall will point in the direction perpendicular to the step
leading to strong demagnetizing fields and therefore to a large
domain wall energy. In a previous paper,\cite{Pennec2004} we have
confirmed that using micromagnetic simulations. Energetically it is
therefore favorable for the domain walls to be situated on the flat
terraces. This is however only possible if the terraces are larger
than the domain wall width. For the smaller miscut angle (\si{}),
the terraces are about 60 nm large and the reversal is found to
occur mainly by nucleation. A typical domain wall width in permalloy
is one hundred nanometers, but some authors have shown that domain
wall width is decreased when the wall is on a step.\cite{Bodea2005}
Also our simulations\cite{Pennec2004} have shown that a domain wall
can be confined on a terrace by a slight compression of its width.
In \si{}, a domain wall can therefore ``fit'' on a terrace and the
energy cost to create a domain wall is relatively small. However, in
order to propagate it will have to overcome a step. This leads to a
stronger domain wall pinning and consequently a magnetization
reversal with mainly nucleation. On the contrary, for samples II and
III the terraces are smaller than the domain wall width in
permalloy. A domain wall in these samples will therefore be partly
located on one or more steps, leading to a higher barrier for domain
nucleation. However, once nucleated the domain walls can propagate
quite easily since the extra energy barrier for passing another step
is rather small. Moreover, the step heights for \sii{} and even more
\siii{} are also smaller than for \si{}. The reversal in these
samples will therefore mainly take place by propagation.

\section{Influence of domain walls}
In the previous section we have shown that the magnetization
reversal in the soft magnetic layer of magnetic tunnel junction-like
trilayers depends on topography. In this section we will show that
also inhomogeneities in the magnetization of the hard layer, like
the presence of magnetic domain walls, can influence the
magnetization reversal of the soft layer. In a previous
paper\cite{Vogel2005}, we have shown that a domain wall in the Co
layer of \sii{} locally decreased the barrier for domain nucleation
in the FeNi layer. We have performed time-resolved XMCD-PEEM
measurements also on the other samples, in order to confirm that the
preferential nucleation induced by domain wall stray fields is a
general property of this type of trilayers. The effect of domain
wall stray fields on the magnetization reversal should be larger for
samples with a higher intrinsic barrier for domain nucleation. This
nucleation barrier strongly depends on topography as shown in the
previous section.

\begin{figure}[ht!]
\includegraphics[width=8.4cm]{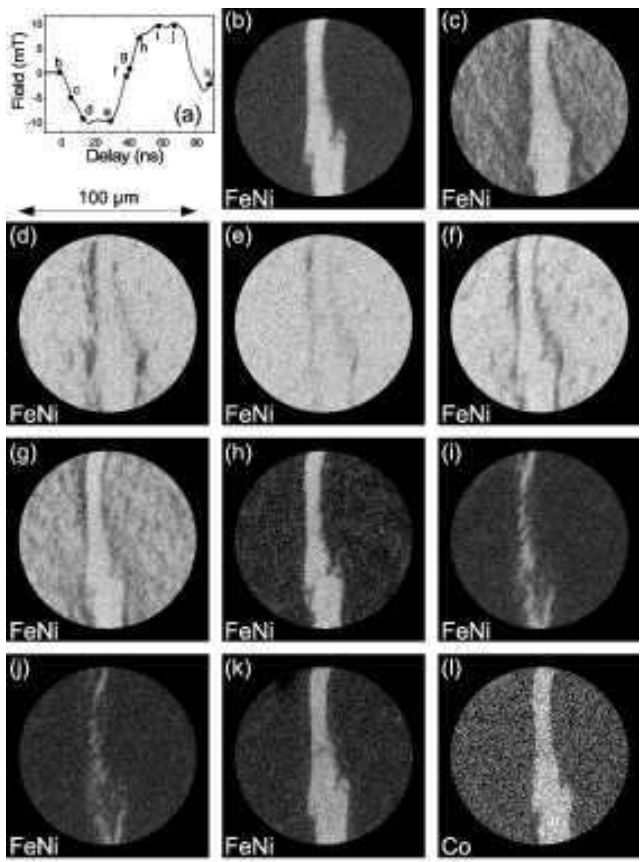}
\caption{\label{fig:PEEM_mis4deg} Time-resolved PEEM images for
\si{} (4$^\circ$ miscut angle). (a) Bipolar applied magnetic field
pulses. (b) to (k) FeNi domain structure for different delays
between magnetic field and photon pulses as indicated in (a). (l) Co
domain structure. The directions of incoming photons, applied field
and local magnetization are the same as in Fig. 4. The field of view
for each image is 100~$\mu$m.}
\end{figure}

\begin{figure}[ht!]
\includegraphics[width=8.4cm]{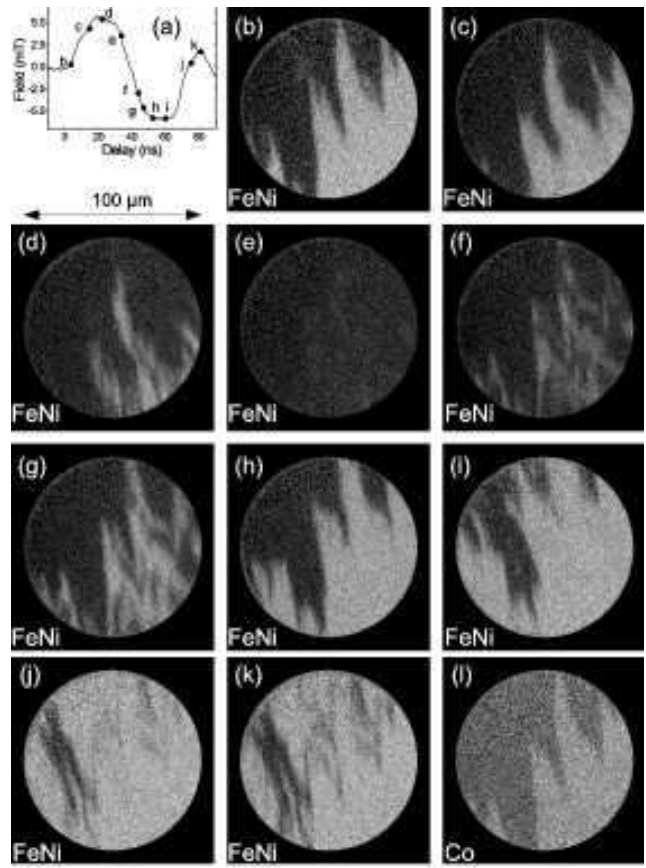}
\caption{\label{fig:PEEM_mis6deg} Time-resolved PEEM images for
\sii{} (6$^\circ$ miscut angle). (a) Bipolar applied magnetic field
pulses. (b) to (k) FeNi domain structure for different delays
between field pulse and photon pulses as indicated in (a). (l) Co
domain structure. The directions of incoming photons, applied field
and local magnetization are the same as in Fig. 4. The field of view
for each image is 100~$\mu$m.}
\end{figure}

\begin{figure}[ht!]
\includegraphics*[bb= 4 523 242 839]{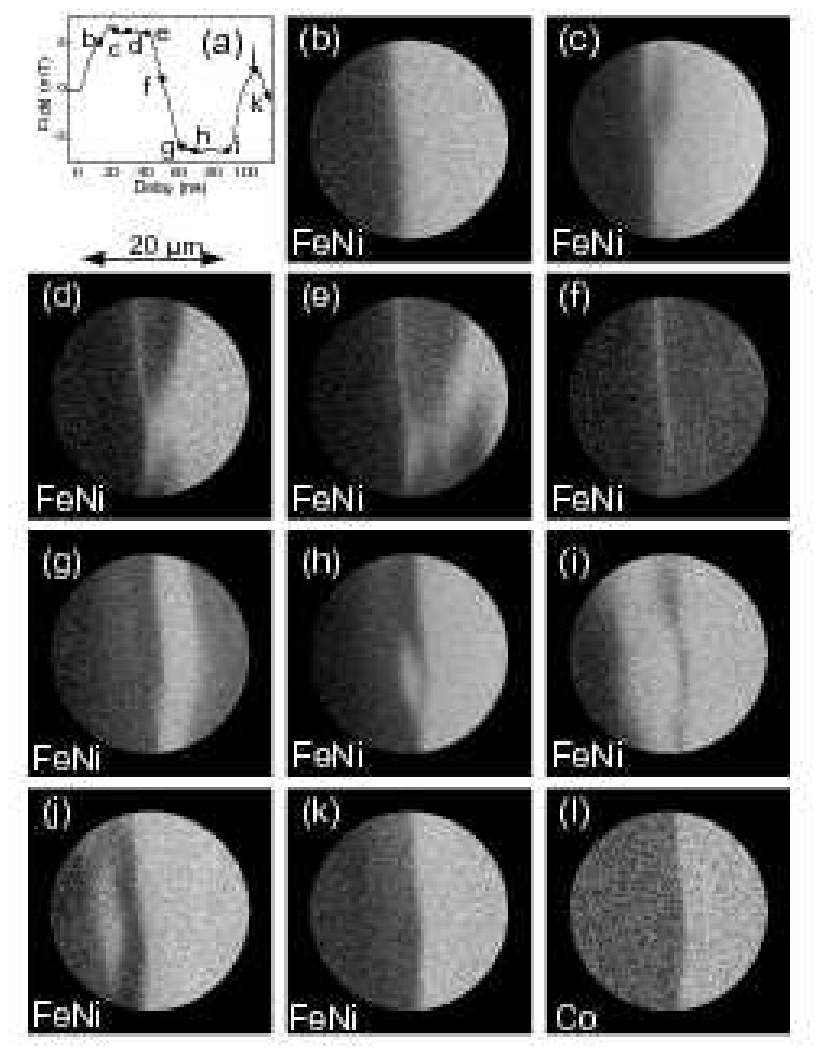}
\caption{\label{fig:PEEM_mis8deg} Time-resolved PEEM images for
\siii{} (8$^\circ$ miscut angle). (a) Bipolar applied magnetic field
pulses. (b) to (k) FeNi domain structure for different delays
between field pulse and photon pulses as indicated in (a). (l) Co
domain structure. The directions of incoming photons, applied field
and local magnetization are the same as in Fig. 4. The field of view
for each image is 20~$\mu$m.}
\end{figure}

The results for the three different samples are shown in
Figs.~\ref{fig:PEEM_mis4deg} to \ref{fig:PEEM_mis8deg}. A domain
structure is initially created in the Co layer by applying a 3~ms
long magnetic pulse of about 3~mT. The domain structure of the Co
layer is shown in
Figs.~\ref{fig:PEEM_mis4deg}(l),~\ref{fig:PEEM_mis6deg}(l)
and~\ref{fig:PEEM_mis8deg}(l) for \si{}, II and III respectively.
Fast bipolar magnetic field pulses, shown in
Figs.~\ref{fig:PEEM_mis4deg}(a),~\ref{fig:PEEM_mis6deg}(a),
and~\ref{fig:PEEM_mis8deg}(a), are then applied to the sample. The
applied field is strong enough to reverse the permalloy
magnetization. Note that even if this field is higher than the
quasistatic cobalt coercive field, the pulse duration is too short
to change the cobalt domain structure (as checked by imaging the Co
magnetization at different times during the magnetic pulse). The
magnetization reversal of the permalloy layer is shown in images (b)
to (k). Secondary electrons generated in the Co layer are attenuated
by the FeNi layer on top, leading to a smaller signal to noise ratio
for images of the Co domain structure.

For \si{}, Fig.~\ref{fig:PEEM_mis4deg}(b) shows that, before the
magnetic field pulse, the FeNi domain pattern is the same as that of
the Co layer. This is a consequence of the ``orange-peel coupling''
which favors parallel alignment of the two layers' magnetizations.
From Figs.~\ref{fig:PEEM_mis4deg}(b) to~\ref{fig:PEEM_mis4deg}(d),
the positive magnetic field reverses the black regions of the FeNi
layer, and the reversal occurs  by nucleation of many white domains.
At the end of the positive field pulse
(Fig.~\ref{fig:PEEM_mis4deg}(e)), the FeNi layer is almost fully
saturated, except a gray region which seems to be correlated to the
position of the Co domain walls. At the beginning of the negative
field pulse, Fig.~\ref{fig:PEEM_mis4deg}(f) shows some fuzzy black
spots which correspond to nucleation of new domains, as well as two
black lines which correspond to domains which were still present at
the end of the first pulse. The FeNi region on top of black Co
domains is the first to be reversed as shown in
Fig.~\ref{fig:PEEM_mis4deg}(g), and this reversal takes place with a
high nucleation density. The negative field pulse is not sufficient
to fully saturate the FeNi layer (Figs.~\ref{fig:PEEM_mis4deg}(i)
and (j)). The field overshoot leads to a FeNi domain structure
similar to that of the Co layer as shown in
Fig.~\ref{fig:PEEM_mis4deg}(k).

Figure~\ref{fig:PEEM_mis6deg} shows the magnetization dynamics of
\sii{}'s permalloy layer. The Co domain structure, which is
unchanged by the magnetic pulse, is shown
in~\ref{fig:PEEM_mis6deg}(l). The reversal of the FeNi layer is more
propagative than for \si{} (Figs.~\ref{fig:PEEM_mis4deg}(c), (d) and
(f) to (i)), confirming the conclusions of the previous section. At
the end of the two magnetic field pulses, the FeNi layer is almost
fully saturated, except for some faint lines at the same position as
the domain walls in Co, which are visible in
Figs.~\ref{fig:PEEM_mis6deg}(f) and (j). One can clearly see in
Figs.~\ref{fig:PEEM_mis6deg}(f) and (k) that the reversal is easier
above the Co domain wall. Reversal is in general favored where the
coupling and the applied field are parallel but the domains which
grow on top of the Co domain wall are larger than elsewhere in the
sample. This means that the reversal on top of the domain walls has
started first. After the negative field pulse, the positive
overshoot make the FeNi domain structure again similar to that of
the Co layer (not shown in Fig.~\ref{fig:PEEM_mis6deg}).

Figure~\ref{fig:PEEM_mis8deg} shows the magnetization dynamics of
\siii{} with a smaller field of view and a better spatial resolution
($\simeq$ 0.3~$\mu$m) than Figs.~\ref{fig:PEEM_mis4deg}
and~\ref{fig:PEEM_mis6deg}. In this series of images, it can be
clearly seen that the region of the FeNi layer above the Co domain
wall is hard to saturate. At the end of the positive field pulse,
Fig.~\ref{fig:PEEM_mis8deg}(f) shows a gray line that corresponds to
the position of the Co domain wall. Also in
Fig.~\ref{fig:PEEM_mis8deg}(i), at the end of the negative pulse, a
gray line is visible at the same position. This means that a
``quasi-wall'' stays present in the FeNi layer even for applied
magnetic fields up to 6~mT. It is interesting to note that the pulse
length of 40 ns is apparently too short to make the Co domain wall
move, even if the field amplitude is well above the Co quasistatic
coercive field. This also means that the effect of strong fields on
the 'quasi-wall' without moving the Co domain wall can only be
studied in these fast dynamic measurements. On the other hand, we
can not completely be sure that the gray zone in Figs.
~\ref{fig:PEEM_mis8deg}(f) and (i) corresponds to a single
quasi-wall and not to an accumulation of several 360$^\circ$ domain
walls with a width smaller that the spatial resolution of the
images. Even more than in the other two samples, the preferential
nucleation above the Co domain wall is particularly clear in
\siii{}. This is due both to the smaller field of view and the
zoom-in on the domain wall and the domination of domain wall
propagation in the reversal of this sample. The difference in energy
barrier for nucleation between the region above the Co domain wall
and other regions in the sample is therefore larger. A sweeping
displacement of the domain walls starting from the Co domain wall is
clearly seen both in Figs.~\ref{fig:PEEM_mis8deg}(c)-(e) for
white-to-black reversal and in Figs.~\ref{fig:PEEM_mis8deg}(h)-(i)
for black-to-white reversal. In Fig.~\ref{fig:PEEM_mis8deg}, we do
not show exactly the same part of the sample in all the images in
order to keep the field of view maximum. This allows a better
visualisation of the sweeping domain wall displacement. We have
checked, however, that the central region in the FeNi images always
corresponds to the domain wall position in the Co layer.

For all three samples we studied, the presence of domain walls in
the hard magnetic layer makes it difficult to saturate the soft
layer. This effect can be understood by taking into account the
domain wall stray field. In our previous paper,\cite{Vogel2005} we
have discussed the effect of this stray field. In particular,
micromagnetic simulations have shown that the Co domain wall stray
field is sufficient to create a region where the FeNi magnetization
is tilted away from the easy axis, creating a so-called quasi-wall.
When applying a magnetic field parallel to the easy axis, the torque
induced by the field on the magnetic moments is higher in this
quasi-wall than elsewhere in the sample, leading to an easier
nucleation above the Co domain walls. This effect is most clearly
visible for \siii{} with the narrowest terraces. In this sample, the
difference between nucleation barriers for domain reversal with and
without domain wall in the Co layer is largest. Moreover, once
nucleated, the domain wall propagation is easiest in this sample. In
\si{}, domain wall propagation is strongly pinned, and even if
preferential nucleation still seems to take place above Co domain
walls, this nucleation does not lead to an easier and faster
switching of the FeNi layer magnetization like in \siii{}.

\section{Conclusion}
We have used time-resolved Kerr effect measurements and
time-resolved XMCD-PEEM imaging to study the nanosecond
layer-resolved magnetization dynamics of FeNi/Al$_2$O$_3$/Co
trilayers deposited on step-bunched Si(111) substrates. We have
revealed a strong dependence of the fast magnetic switching on the
layer topography. The phenomenon dominating magnetic switching
(domain nucleation or domain wall propagation) strongly depends on
the average width of the substrate terraces and the height of the
steps between terraces. For relatively large terraces of 60~nm
separated by 5~nm high steps, domain nucleation is dominating while
domain propagation is strongly pinned. On the other hand, for
terraces with a width of only 20~nm separated by 3~nm high steps,
domain wall propagation dominates over domain nucleation. We propose
that this difference can be explained by taking into account the
domain wall width. While a (compressed) domain wall can fit a single
60~nm terrace, it will extend over one or more steps for 20 nm~wide
terraces. The minimum domain wall energy will therefore be smaller
for 60 nm wide terraces, but the energy barrier for crossing a step
will be larger. Coercivity measurements as a function of applied
magnetic field sweep rate and magnetization relaxation measurements
can give an indication of the magnetization reversal mechanisms and
their evolution as a function of time and applied field. However,
our results clearly show that for a comprehensive picture of fast
magnetic switching a microscopic technique is needed. Using the
element selectivity of time-resolved XMCD-PEEM, we have revealed the
influence of stray fields emerging from domain walls in the hard
magnetic layer on the magnetic switching of the soft magnetic layer.
These stray fields locally decrease the barrier for domain
nucleation and lead to a higher switching speed. This effect is
largest for samples where the overall nucleation barrier is high and
where reversal takes place mainly by domain wall propagation. The
influence of domain wall stray fields can therefore be modified by
changing the substrate topography, and can be used to manipulate
local switching speeds and switching reproducibility.

We thank A. Vaur\`{e}s for sample preparation. Financial support by
EU (BESSY-EC-HPRI contract No. HPRI-1999-CT-00028) and the
Laboratoire Europ\'{e}en Associ\'{e} ``Mesomag'' is gratefully
acknowledged. J.C., F.R. and J.V. acknowledge partial financial
support for personnel exchange by the ``Acciones
Integradas-Picasso'' Programme, through Grant No. HF2003-0173.


\begin{thebibliography}{22}
\expandafter\ifx\csname natexlab\endcsname\relax\def\natexlab#1{#1}\fi
\expandafter\ifx\csname bibnamefont\endcsname\relax
  \def\bibnamefont#1{#1}\fi
\expandafter\ifx\csname bibfnamefont\endcsname\relax
  \def\bibfnamefont#1{#1}\fi
\expandafter\ifx\csname citenamefont\endcsname\relax
  \def\citenamefont#1{#1}\fi
\expandafter\ifx\csname url\endcsname\relax
  \def\url#1{\texttt{#1}}\fi
\expandafter\ifx\csname urlprefix\endcsname\relax\def\urlprefix{URL }\fi
\providecommand{\bibinfo}[2]{#2}
\providecommand{\eprint}[2][]{\url{#2}}

\bibitem[{\citenamefont{Baibich et~al.}(1988)\citenamefont{Baibich, Broto,
  Fert, Dau, Petroff, Etienne, Creuzet, Friederich, and
  Chazelas}}]{Baibich1988}
\bibinfo{author}{\bibfnamefont{M.~N.} \bibnamefont{Baibich}},
  \bibinfo{author}{\bibfnamefont{J.~M.} \bibnamefont{Broto}},
  \bibinfo{author}{\bibfnamefont{A.}~\bibnamefont{Fert}},
  \bibinfo{author}{\bibfnamefont{F.}~\bibnamefont{Nguyen Van Dau}},
  \bibinfo{author}{\bibfnamefont{F.}~\bibnamefont{Petroff}},
  \bibinfo{author}{\bibfnamefont{P.}~\bibnamefont{Etienne}},
  \bibinfo{author}{\bibfnamefont{G.}~\bibnamefont{Creuzet}},
  \bibinfo{author}{\bibfnamefont{A.}~\bibnamefont{Friederich}},
  \bibnamefont{and} \bibinfo{author}{\bibfnamefont{J.}~\bibnamefont{Chazelas}},
  \bibinfo{journal}{Phys. Rev. Lett.} \textbf{\bibinfo{volume}{61}},
  \bibinfo{pages}{2472} (\bibinfo{year}{1988}).

\bibitem[{\citenamefont{Binasch et~al.}(1989)\citenamefont{Binasch, Grünberg,
  Saurenbach, and Zinn}}]{Binasch1989}
\bibinfo{author}{\bibfnamefont{G.}~\bibnamefont{Binasch}},
  \bibinfo{author}{\bibfnamefont{P.}~\bibnamefont{Gr\"{u}nberg}},
  \bibinfo{author}{\bibfnamefont{F.}~\bibnamefont{Saurenbach}},
  \bibnamefont{and} \bibinfo{author}{\bibfnamefont{W.}~\bibnamefont{Zinn}},
  \bibinfo{journal}{Phys. Rev. B} \textbf{\bibinfo{volume}{39}},
  \bibinfo{pages}{4828} (\bibinfo{year}{1989}).

\bibitem[{\citenamefont{Myers et~al.}(1999)\citenamefont{Myers, Ralph, Katine,
  Louie, and Buhrman}}]{Myers1999}
\bibinfo{author}{\bibfnamefont{E.~B.} \bibnamefont{Myers}},
  \bibinfo{author}{\bibfnamefont{D.~C.} \bibnamefont{Ralph}},
  \bibinfo{author}{\bibfnamefont{J.~A.} \bibnamefont{Katine}},
  \bibinfo{author}{\bibfnamefont{R.~N.} \bibnamefont{Louie}}, \bibnamefont{and}
  \bibinfo{author}{\bibfnamefont{R.~A.} \bibnamefont{Buhrman}},
  \bibinfo{journal}{Science} \textbf{\bibinfo{volume}{285}},
  \bibinfo{pages}{867} (\bibinfo{year}{1999}).

\bibitem[{\citenamefont{Weber et~al.}(2001)\citenamefont{Weber, Riesen, and
  Siegmann}}]{Weber2001}
\bibinfo{author}{\bibfnamefont{W.}~\bibnamefont{Weber}},
  \bibinfo{author}{\bibfnamefont{S.}~\bibnamefont{Riesen}}, \bibnamefont{and}
  \bibinfo{author}{\bibfnamefont{H.~C.} \bibnamefont{Siegmann}},
  \bibinfo{journal}{Science} \textbf{\bibinfo{volume}{291}},
  \bibinfo{pages}{1015} (\bibinfo{year}{2001}).

\bibitem[{\citenamefont{N\'{e}el}(1962)}]{Neel1962}
\bibinfo{author}{\bibfnamefont{L.}~\bibnamefont{N\'{e}el}}, \bibinfo{journal}{C. R.
  Hebd. Seances Acad. Sci.} \textbf{\bibinfo{volume}{255}},
  \bibinfo{pages}{1676} (\bibinfo{year}{1962}).

\bibitem[{\citenamefont{Pennec et~al.}(2004)\citenamefont{Pennec, Camarero,
  Toussaint, Pizzini, Bonfim, Petroff, Kuch, Offi, Fukumoto, {Nguyen Van Dau}
  et~al.}}]{Pennec2004}
\bibinfo{author}{\bibfnamefont{Y.}~\bibnamefont{Pennec}},
  \bibinfo{author}{\bibfnamefont{J.}~\bibnamefont{Camarero}},
  \bibinfo{author}{\bibfnamefont{J.~C.} \bibnamefont{Toussaint}},
  \bibinfo{author}{\bibfnamefont{S.}~\bibnamefont{Pizzini}},
  \bibinfo{author}{\bibfnamefont{M.}~\bibnamefont{Bonfim}},
  \bibinfo{author}{\bibfnamefont{F.}~\bibnamefont{Petroff}},
  \bibinfo{author}{\bibfnamefont{W.}~\bibnamefont{Kuch}},
  \bibinfo{author}{\bibfnamefont{F.}~\bibnamefont{Offi}},
  \bibinfo{author}{\bibfnamefont{K.}~\bibnamefont{Fukumoto}},
  \bibinfo{author}{\bibfnamefont{F.}~\bibnamefont{{Nguyen Van Dau}}},
  \bibnamefont{and}
  \bibinfo{author}{\bibfnamefont{J.}~\bibnamefont{Vogel}},
  \bibinfo{journal}{Phys. Rev. B}
  \textbf{\bibinfo{volume}{69}}, \bibinfo{pages}{180402(R)}
  (\bibinfo{year}{2004}).

\bibitem[{\citenamefont{Thomas et~al.}(2000)\citenamefont{Thomas, Samant, and
  Parkin}}]{Thomas2000}
\bibinfo{author}{\bibfnamefont{L.}~\bibnamefont{Thomas}},
  \bibinfo{author}{\bibfnamefont{M.~G.} \bibnamefont{Samant}},
  \bibnamefont{and} \bibinfo{author}{\bibfnamefont{S.~S.~P.}
  \bibnamefont{Parkin}}, \bibinfo{journal}{Phys. Rev. Lett.}
  \textbf{\bibinfo{volume}{84}}, \bibinfo{pages}{1816} (\bibinfo{year}{2000}).

\bibitem[{\citenamefont{Sch\"afer et~al.}(2002)\citenamefont{Sch\"afer, Urban,
  Ullmann, Meyerheim, Heinrich, Schultz, and Kirschner}}]{Schafer2002}
\bibinfo{author}{\bibfnamefont{R.}~\bibnamefont{Sch\"afer}},
  \bibinfo{author}{\bibfnamefont{R.}~\bibnamefont{Urban}},
  \bibinfo{author}{\bibfnamefont{D.}~\bibnamefont{Ullmann}},
  \bibinfo{author}{\bibfnamefont{H.~L.} \bibnamefont{Meyerheim}},
  \bibinfo{author}{\bibfnamefont{B.}~\bibnamefont{Heinrich}},
  \bibinfo{author}{\bibfnamefont{L.}~\bibnamefont{Schultz}}, \bibnamefont{and}
  \bibinfo{author}{\bibfnamefont{J.}~\bibnamefont{Kirschner}},
  \bibinfo{journal}{Phys. Rev. B} \textbf{\bibinfo{volume}{65}},
  \bibinfo{pages}{144405} (\bibinfo{year}{2002}).

\bibitem[{\citenamefont{Kuch et~al.}(2003)\citenamefont{Kuch, Chelaru,
  Fukumoto, Porrati, Offi, Kotsugi, and Kirschner}}]{Kuch2003}
\bibinfo{author}{\bibfnamefont{W.}~\bibnamefont{Kuch}},
  \bibinfo{author}{\bibfnamefont{L.~I.} \bibnamefont{Chelaru}},
  \bibinfo{author}{\bibfnamefont{K.}~\bibnamefont{Fukumoto}},
  \bibinfo{author}{\bibfnamefont{F.}~\bibnamefont{Porrati}},
  \bibinfo{author}{\bibfnamefont{F.}~\bibnamefont{Offi}},
  \bibinfo{author}{\bibfnamefont{M.}~\bibnamefont{Kotsugi}}, \bibnamefont{and}
  \bibinfo{author}{\bibfnamefont{J.}~\bibnamefont{Kirschner}},
  \bibinfo{journal}{Phys. Rev. B} \textbf{\bibinfo{volume}{67}},
  \bibinfo{pages}{214403} (\bibinfo{year}{2003}).

\bibitem[{\citenamefont{Christoph and Sch\"{a}fer}(2004)}]{Christoph2004}
\bibinfo{author}{\bibfnamefont{V.}~\bibnamefont{Christoph}} \bibnamefont{and}
  \bibinfo{author}{\bibfnamefont{R.}~\bibnamefont{Sch\"afer}},
  \bibinfo{journal}{Phys. Rev. B} \textbf{\bibinfo{volume}{70}}
  \bibinfo{pages}{214419} (\bibinfo{year}{2004}).

\bibitem[{\citenamefont{Wiebel et~al.}(2005)\citenamefont{Wiebel, Jamet,
  Vernier, Mougin, Ferré, Baltz, Rodmacq, and Dieny}}]{Wiebel2005}
\bibinfo{author}{\bibfnamefont{S.}~\bibnamefont{Wiebel}},
  \bibinfo{author}{\bibfnamefont{J.-P.} \bibnamefont{Jamet}},
  \bibinfo{author}{\bibfnamefont{N.}~\bibnamefont{Vernier}},
  \bibinfo{author}{\bibfnamefont{A.}~\bibnamefont{Mougin}},
  \bibinfo{author}{\bibfnamefont{J.}~\bibnamefont{Ferré}},
  \bibinfo{author}{\bibfnamefont{V.}~\bibnamefont{Baltz}},
  \bibinfo{author}{\bibfnamefont{B.}~\bibnamefont{Rodmacq}}, \bibnamefont{and}
  \bibinfo{author}{\bibfnamefont{B.}~\bibnamefont{Dieny}},
  \bibinfo{journal}{Appl. Phys. Lett.} \textbf{\bibinfo{volume}{86}},
  \bibinfo{pages}{142502} (\bibinfo{year}{2005}).

\bibitem[{\citenamefont{Vogel et~al.}(2005)\citenamefont{Vogel, Kuch, Hertel,
  Camarero, Fukumoto, Romanens, Pizzini, Bonfim, Petroff, Fontaine
  et~al.}}]{Vogel2005}
\bibinfo{author}{\bibfnamefont{J.}~\bibnamefont{Vogel}},
  \bibinfo{author}{\bibfnamefont{W.}~\bibnamefont{Kuch}},
  \bibinfo{author}{\bibfnamefont{R.}~\bibnamefont{Hertel}},
  \bibinfo{author}{\bibfnamefont{J.}~\bibnamefont{Camarero}},
  \bibinfo{author}{\bibfnamefont{K.}~\bibnamefont{Fukumoto}},
  \bibinfo{author}{\bibfnamefont{F.}~\bibnamefont{Romanens}},
  \bibinfo{author}{\bibfnamefont{S.}~\bibnamefont{Pizzini}},
  \bibinfo{author}{\bibfnamefont{M.}~\bibnamefont{Bonfim}},
  \bibinfo{author}{\bibfnamefont{F.}~\bibnamefont{Petroff}},
  \bibinfo{author}{\bibfnamefont{A.}~\bibnamefont{Fontaine}},
  \bibnamefont{and}
  \bibinfo{author}{\bibfnamefont{J.}~\bibnamefont{Kirschner}},
  \bibinfo{journal}{Phys. Rev. B}
  \textbf{\bibinfo{volume}{72}}, \bibinfo{pages}{220402(R)}
  (\bibinfo{year}{2005}).

\bibitem[{\citenamefont{Sussiau}(1997)}]{SussiauPhD}
\bibinfo{author}{\bibfnamefont{M.}~\bibnamefont{Sussiau}}, Ph.D. thesis,
  \bibinfo{school}{Université Paris Sud} (\bibinfo{year}{1997}).

\bibitem[{\citenamefont{Vogel et~al.}(2004)\citenamefont{Vogel, Kuch, Camarero,
  Fukumoto, Pennec, Bonfim, Pizzini, Petroff, Fontaine, and
  Kirschner}}]{Vogel2004}
\bibinfo{author}{\bibfnamefont{J.}~\bibnamefont{Vogel}},
  \bibinfo{author}{\bibfnamefont{W.}~\bibnamefont{Kuch}},
  \bibinfo{author}{\bibfnamefont{J.}~\bibnamefont{Camarero}},
  \bibinfo{author}{\bibfnamefont{K.}~\bibnamefont{Fukumoto}},
  \bibinfo{author}{\bibfnamefont{Y.}~\bibnamefont{Pennec}},
  \bibinfo{author}{\bibfnamefont{M.}~\bibnamefont{Bonfim}},
  \bibinfo{author}{\bibfnamefont{S.}~\bibnamefont{Pizzini}},
  \bibinfo{author}{\bibfnamefont{F.}~\bibnamefont{Petroff}},
  \bibinfo{author}{\bibfnamefont{A.}~\bibnamefont{Fontaine}}, \bibnamefont{and}
  \bibinfo{author}{\bibfnamefont{J.}~\bibnamefont{Kirschner}},
  \bibinfo{journal}{J. Appl. Phys.} \textbf{\bibinfo{volume}{95}},
  \bibinfo{pages}{6533} (\bibinfo{year}{2004}).

\bibitem[{\citenamefont{Stohr}(1999)\citenamefont{Stohr}}]{Stohr1999}
\bibinfo{author}{\bibfnamefont{J.}~\bibnamefont{St\"{o}hr}},
\bibinfo{journal}{J. Magn. Magn. Mater.} \textbf{\bibinfo{volume}{200}},
\bibinfo{pages}{470} (\bibinfo{year}{1999}).

\bibitem[{\citenamefont{Wende}(2004)\citenamefont{Wende}}]{Wende2004}
\bibinfo{author}{\bibfnamefont{H.}~\bibnamefont{Wende}},
\bibinfo{journal}{Rep. Prog. Phys.} \textbf{\bibinfo{volume}{67}},
\bibinfo{pages}{2105} (\bibinfo{year}{2004}).

\bibitem[{\citenamefont{Schneider and Schonhense}(2002)\citenamefont{Schneider}}]{Schneider2002}
\bibinfo{author}{\bibfnamefont{C.M.}~\bibnamefont{Schneider}},
\bibinfo{author}{\bibfnamefont{G.}~\bibnamefont{Sch\"{o}nhense}},
\bibinfo{journal}{Rep. Prog. Phys.} \textbf{\bibinfo{volume}{65}}, \bibinfo{pages}{1785} (\bibinfo{year}{2002}).

\bibitem[{\citenamefont{Kuch}(2004)\citenamefont{Kuch}}]{Kuch2004}
\bibinfo{author}{\bibfnamefont{W.}~\bibnamefont{Kuch}},
\bibinfo{journal}{Phys. Scr.} \textbf{\bibinfo{volume}{T109}}, \bibinfo{pages}{89} (\bibinfo{year}{2004}).

\bibitem[{\citenamefont{Bonfim et~al.}(2001)\citenamefont{Bonfim, Ghiringhelli,
Montaigne, Pizzini, Brookes, Petroff, Vogel, Camarero, and
Fontaine}}]{Bonfim2001}
\bibinfo{author}{\bibfnamefont{M.}~\bibnamefont{Bonfim}},
  \bibinfo{author}{\bibfnamefont{G.}~\bibnamefont{Ghiringhelli}},
  \bibinfo{author}{\bibfnamefont{F.}~\bibnamefont{Montaigne}},
  \bibinfo{author}{\bibfnamefont{S.}~\bibnamefont{Pizzini}},
  \bibinfo{author}{\bibfnamefont{N.B.}~\bibnamefont{Brookes}},
  \bibinfo{author}{\bibfnamefont{F.}~\bibnamefont{Petroff}},
  \bibinfo{author}{\bibfnamefont{J.}~\bibnamefont{Vogel}},
  \bibinfo{author}{\bibfnamefont{J.}~\bibnamefont{Camarero}}, \bibnamefont{and}
  \bibinfo{author}{\bibfnamefont{A.}~\bibnamefont{Fontaine}},
  \bibinfo{journal}{Phys. Rev. Lett.} \textbf{\bibinfo{volume}{86}},
  \bibinfo{pages}{3646} (\bibinfo{year}{2001}).

\bibitem[{\citenamefont{Vogel et~al.}(2003)\citenamefont{Vogel, Kuch, Bonfim,
  Camarero, Pennec, Offi, Fukumoto, Kirschner, Fontaine, and
  Pizzini}}]{Vogel2003}
\bibinfo{author}{\bibfnamefont{J.}~\bibnamefont{Vogel}},
  \bibinfo{author}{\bibfnamefont{W.}~\bibnamefont{Kuch}},
  \bibinfo{author}{\bibfnamefont{M.}~\bibnamefont{Bonfim}},
  \bibinfo{author}{\bibfnamefont{J.}~\bibnamefont{Camarero}},
  \bibinfo{author}{\bibfnamefont{Y.}~\bibnamefont{Pennec}},
  \bibinfo{author}{\bibfnamefont{F.}~\bibnamefont{Offi}},
  \bibinfo{author}{\bibfnamefont{K.}~\bibnamefont{Fukumoto}},
  \bibinfo{author}{\bibfnamefont{J.}~\bibnamefont{Kirschner}},
  \bibinfo{author}{\bibfnamefont{A.}~\bibnamefont{Fontaine}}, \bibnamefont{and}
  \bibinfo{author}{\bibfnamefont{S.}~\bibnamefont{Pizzini}},
  \bibinfo{journal}{Appl. Phys. Lett.} \textbf{\bibinfo{volume}{82}},
  \bibinfo{pages}{2299} (\bibinfo{year}{2003}).

\bibitem[{\citenamefont{Raquet et~al.}(1995)\citenamefont{Raquet, Ortega,
  Goiran, Fert, Redoules, Mamy, Ousset, Sdaq, and Khmou}}]{Raquet1995}
\bibinfo{author}{\bibfnamefont{B.}~\bibnamefont{Raquet}},
  \bibinfo{author}{\bibfnamefont{M.}~\bibnamefont{Ortega}},
  \bibinfo{author}{\bibfnamefont{M.}~\bibnamefont{Goiran}},
  \bibinfo{author}{\bibfnamefont{A.~R.} \bibnamefont{Fert}},
  \bibinfo{author}{\bibfnamefont{J.~P.} \bibnamefont{Redoules}},
  \bibinfo{author}{\bibfnamefont{R.}~\bibnamefont{Mamy}},
  \bibinfo{author}{\bibfnamefont{J.~C.} \bibnamefont{Ousset}},
  \bibinfo{author}{\bibfnamefont{A.}~\bibnamefont{Sdaq}}, \bibnamefont{and}
  \bibinfo{author}{\bibfnamefont{A.}~\bibnamefont{Khmou}}, \bibinfo{journal}{J.
  Magn. Magn. Mater.} \textbf{\bibinfo{volume}{150}}, \bibinfo{pages}{L5}
  (\bibinfo{year}{1995}).

\bibitem[{\citenamefont{Camarero et~al.}(2001)\citenamefont{Camarero, Pennec,
  Vogel, Bonfim, Pizzini, Cartier, Ernult, Fettar, and Dieny}}]{Camarero2001}
\bibinfo{author}{\bibfnamefont{J.}~\bibnamefont{Camarero}},
  \bibinfo{author}{\bibfnamefont{Y.}~\bibnamefont{Pennec}},
  \bibinfo{author}{\bibfnamefont{J.}~\bibnamefont{Vogel}},
  \bibinfo{author}{\bibfnamefont{M.}~\bibnamefont{Bonfim}},
  \bibinfo{author}{\bibfnamefont{S.}~\bibnamefont{Pizzini}},
  \bibinfo{author}{\bibfnamefont{M.}~\bibnamefont{Cartier}},
  \bibinfo{author}{\bibfnamefont{F.}~\bibnamefont{Ernult}},
  \bibinfo{author}{\bibfnamefont{F.}~\bibnamefont{Fettar}}, \bibnamefont{and}
  \bibinfo{author}{\bibfnamefont{B.}~\bibnamefont{Dieny}},
  \bibinfo{journal}{Phys. Rev. B} \textbf{\bibinfo{volume}{64}},
  \bibinfo{pages}{172402} (\bibinfo{year}{2001}).

\bibitem[{\citenamefont{Raquet et~al.}(1996)\citenamefont{Raquet, Mamy, and
  Ousset}}]{Raquet1996}
\bibinfo{author}{\bibfnamefont{B.}~\bibnamefont{Raquet}},
  \bibinfo{author}{\bibfnamefont{R.}~\bibnamefont{Mamy}}, \bibnamefont{and}
  \bibinfo{author}{\bibfnamefont{J.~C.} \bibnamefont{Ousset}},
  \bibinfo{journal}{Phys. Rev. B} \textbf{\bibinfo{volume}{54}},
  \bibinfo{pages}{4128} (\bibinfo{year}{1996}).

\bibitem[{\citenamefont{Romanens et~al.}(2005)\citenamefont{Romanens, Pizzini,
  Sort, Garcia, Camarero, Yokaichiya, Pennec, Vogel, and
  Dieny}}]{Romanens2005a}
\bibinfo{author}{\bibfnamefont{F.}~\bibnamefont{Romanens}},
  \bibinfo{author}{\bibfnamefont{S.}~\bibnamefont{Pizzini}},
  \bibinfo{author}{\bibfnamefont{J.}~\bibnamefont{Sort}},
  \bibinfo{author}{\bibfnamefont{F.}~\bibnamefont{Garcia}},
  \bibinfo{author}{\bibfnamefont{J.}~\bibnamefont{Camarero}},
  \bibinfo{author}{\bibfnamefont{F.}~\bibnamefont{Yokaichiya}},
  \bibinfo{author}{\bibfnamefont{Y.}~\bibnamefont{Pennec}},
  \bibinfo{author}{\bibfnamefont{J.}~\bibnamefont{Vogel}}, \bibnamefont{and}
  \bibinfo{author}{\bibfnamefont{B.}~\bibnamefont{Dieny}},
  \bibinfo{journal}{Eur. Phys. J. B} \textbf{\bibinfo{volume}{45}},
  \bibinfo{pages}{185} (\bibinfo{year}{2005}).

\bibitem[{\citenamefont{Moritz et~al.}(2004)\citenamefont{Moritz, Dieny, Nozi\`{e}res,
Pennec, Camarero, and Pizzini}}]{Moritz2004}
\bibinfo{author}{\bibfnamefont{J.}~\bibnamefont{Moritz}},
  \bibinfo{author}{\bibfnamefont{B.}~\bibnamefont{Dieny}},
  \bibinfo{author}{\bibfnamefont{J.P.}~\bibnamefont{Nozi\`{e}res}},
  \bibinfo{author}{\bibfnamefont{Y.}~\bibnamefont{Pennec}},
  \bibinfo{author}{\bibfnamefont{J.}~\bibnamefont{Camarero}}, \bibnamefont{and}
  \bibinfo{author}{\bibfnamefont{S.}~\bibnamefont{Pizzini}},
  \bibinfo{journal}{Phys. Rev. B} \textbf{\bibinfo{volume}{71}},
  \bibinfo{pages}{100402(R)} (\bibinfo{year}{2004}).

\bibitem[{\citenamefont{Schumacher et~al.}(2003)\citenamefont{Schumacher, Chappert,
Sousa, Freitas, and Miltat}}]{Schumacher2003}
\bibinfo{author}{\bibfnamefont{H.W.}~\bibnamefont{Schumacher}},
  \bibinfo{author}{\bibfnamefont{C.}~\bibnamefont{Chappert}},
  \bibinfo{author}{\bibfnamefont{R.C.}~\bibnamefont{Sousa}},
  \bibinfo{author}{\bibfnamefont{P.P.}~\bibnamefont{Freitas}}, \bibnamefont{and}
  \bibinfo{author}{\bibfnamefont{J.}~\bibnamefont{Miltat}},
  \bibinfo{journal}{Phys. Rev. Lett.} \textbf{\bibinfo{volume}{90}},
  \bibinfo{pages}{017204} (\bibinfo{year}{2003}).

\bibitem[{\citenamefont{Bailey et~al.}(2004)\citenamefont{Bailey, Cheng, Keavney,
Kao, Vescovo, and Arena}}]{Bailey2004}
\bibinfo{author}{\bibfnamefont{W.E.}~\bibnamefont{Bailey}},
  \bibinfo{author}{\bibfnamefont{L.}~\bibnamefont{Cheng}},
  \bibinfo{author}{\bibfnamefont{D.J.}~\bibnamefont{Keavney}},
  \bibinfo{author}{\bibfnamefont{C.-C.}~\bibnamefont{Kao}},
  \bibinfo{author}{\bibfnamefont{E.}~\bibnamefont{Vescovo}}, \bibnamefont{and}
  \bibinfo{author}{\bibfnamefont{D.A.}~\bibnamefont{Arena}},
  \bibinfo{journal}{Phys. Rev. B} \textbf{\bibinfo{volume}{70}},
  \bibinfo{pages}{172403} (\bibinfo{year}{2004}).

\bibitem[{\citenamefont{Choi et~al.}(2005)\citenamefont{Choi, Ho, Arnup,
and Freeman}}]{Choi2005}
\bibinfo{author}{\bibfnamefont{B.C.}~\bibnamefont{Choi}},
  \bibinfo{author}{\bibfnamefont{J.}~\bibnamefont{Ho}},
  \bibinfo{author}{\bibfnamefont{G.}~\bibnamefont{Arnup}}, \bibnamefont{and}
  \bibinfo{author}{\bibfnamefont{M.R.}~\bibnamefont{Freeman}},
  \bibinfo{journal}{Phys. Rev. Lett.} \textbf{\bibinfo{volume}{95}},
  \bibinfo{pages}{237211} (\bibinfo{year}{2005}).

\bibitem[{\citenamefont{Krasyuk et~al.}(2005)\citenamefont{Krasyuk, Wegelin,
Nepijko, Elmers, Sch\"{o}nhense, Bolte, and
Schneider}}]{Krasyuk2005}
\bibinfo{author}{\bibfnamefont{A.}~\bibnamefont{Krasyuk}},
  \bibinfo{author}{\bibfnamefont{F.}~\bibnamefont{Wegelin}},
  \bibinfo{author}{\bibfnamefont{S.A.}~\bibnamefont{Nepijko}},
  \bibinfo{author}{\bibfnamefont{H.J.}~\bibnamefont{Elmers}},
\bibinfo{author}{\bibfnamefont{G.}~\bibnamefont{Sch\"{o}nhense}},
  \bibinfo{author}{\bibfnamefont{M.}~\bibnamefont{Bolte}}, \bibnamefont{and}
  \bibinfo{author}{\bibfnamefont{C.M.}~\bibnamefont{Schneider}},
  \bibinfo{journal}{Phys. Rev. Lett.} \textbf{\bibinfo{volume}{95}},
  \bibinfo{pages}{207201} (\bibinfo{year}{2005}).

\bibitem[{\citenamefont{Fatuzzo}(1962)}]{Fatuzzo1962}
\bibinfo{author}{\bibfnamefont{E.}~\bibnamefont{Fatuzzo}},
  \bibinfo{journal}{Phys. Rev.} \textbf{\bibinfo{volume}{127}},
  \bibinfo{pages}{1999} (\bibinfo{year}{1962}).

\bibitem[{\citenamefont{Labrune et~al.}(1989)\citenamefont{Labrune, Andrieu,
  Rio, and Bernstein}}]{Labrune1989}
\bibinfo{author}{\bibfnamefont{M.}~\bibnamefont{Labrune}},
  \bibinfo{author}{\bibfnamefont{S.}~\bibnamefont{Andrieu}},
  \bibinfo{author}{\bibfnamefont{F.}~\bibnamefont{Rio}}, \bibnamefont{and}
  \bibinfo{author}{\bibfnamefont{P.}~\bibnamefont{Bernstein}},
  \bibinfo{journal}{J. Magn. Magn. Mater.} \textbf{\bibinfo{volume}{80}},
  \bibinfo{pages}{211} (\bibinfo{year}{1989}).

\bibitem[{\citenamefont{Fukumoto et~al.}(2006)\citenamefont{Fukumoto, Kuch,
Vogel, Romanens, Pizzini, Camarero, Bonfim, and
Kirschner}}]{Fukumoto2006}
\bibinfo{author}{\bibfnamefont{K.}~\bibnamefont{Fukumoto}},
  \bibinfo{author}{\bibfnamefont{W.}~\bibnamefont{Kuch}},
  \bibinfo{author}{\bibfnamefont{J.}~\bibnamefont{Vogel}},
  \bibinfo{author}{\bibfnamefont{F.}~\bibnamefont{Romanens}},
  \bibinfo{author}{\bibfnamefont{S.}~\bibnamefont{Pizzini}},
  \bibinfo{author}{\bibfnamefont{J.}~\bibnamefont{Camarero}},
  \bibinfo{author}{\bibfnamefont{M.}~\bibnamefont{Bonfim}},
  \bibnamefont{and}
  \bibinfo{author}{\bibfnamefont{J.}~\bibnamefont{Kirschner}},
  \bibinfo{journal}{Phys. Rev. Lett.}
  \textbf{\bibinfo{volume}{96}}, \bibinfo{pages}{097204}
  (\bibinfo{year}{2006}).

\bibitem[{\citenamefont{Hubert and Sch\"{a}fer}(1998)\citenamefont{Hubert
and Sch\"{a}fer}}]{Hubertbook}
\bibinfo{author}{\bibfnamefont{A.}~\bibnamefont{Hubert}} \bibnamefont{and}
  \bibinfo{author}{\bibfnamefont{R.}~\bibnamefont{Sch\"{a}fer}},
  \textit{\bibinfo{book}{Magnetic Domains: The Analysis of Magnetic
  Microstructures}}, \bibinfo{publisher}{Springer-Verlag, Berlin}
  (\bibinfo{year}{1998}).

\bibitem[{\citenamefont{Bodea et~al.}(2005)\citenamefont{Bodea, Wulfhekel, and
  Kirschner}}]{Bodea2005}
\bibinfo{author}{\bibfnamefont{S.}~\bibnamefont{Bodea}},
  \bibinfo{author}{\bibfnamefont{W.}~\bibnamefont{Wulfhekel}},
  \bibnamefont{and}
  \bibinfo{author}{\bibfnamefont{J.}~\bibnamefont{Kirschner}},
  \bibinfo{journal}{Phys. Rev. B} \textbf{\bibinfo{volume}{72}},
  \bibinfo{pages}{100403(R)} (\bibinfo{year}{2005}).

\end{thebibliography}
\end{document}